\documentclass{article}%
\usepackage{amsmath}
\usepackage{amsfonts}
\usepackage{amssymb}
\usepackage{graphicx}
\usepackage{setspace}%
\providecommand{\U}[1]{\protect\rule{.1in}{.1in}}
\newtheorem{theorem}{Theorem} [section]

\newtheorem{corollary}[theorem]{Corollary}

\newtheorem{definition}[theorem]{Definition}

\newtheorem{lemma}[theorem]{Lemma}

\newtheorem{problem}[theorem]{Problem}

\newenvironment{proof}[1][Proof]{\noindent\textbf{#1.} }{\ \rule{0.5em}{0.5em}}
\begin{document}

\author{Vadim E. Levit\\Ariel University Center of Samaria, ISRAEL\\E-mail: levitv@ariel.ac.il
\and Eugen Mandrescu\\Holon Institute of Technology, ISRAEL\\E-mail: eugen\_m@hit.ac.il }
\date{}
\title{Local Maximum Stable Sets Greedoids Stemmed from Very Well-Covered Graphs}
\maketitle

\begin{abstract}
A \textit{maximum stable set }in a graph $G$ is a stable set of maximum
cardinality. $S$ is called a \textit{local maximum stable set} of $G$, and we
write $S\in\Psi(G)$, if $S$ is a maximum stable set of the subgraph induced by
the closed neighborhood of $S$. A greedoid $(V,\mathcal{F})$ is called a
\textit{local maximum stable set greedoid} if there exists a graph $G=(V,E)$
such that $\mathcal{F}=\Psi(G)$.

Nemhauser and Trotter Jr. \cite{NemhTro}, proved that any $S\in\Psi(G)$ is a
subset of a maximum stable set of $G$. In \cite{LevMan2} we have shown that
the family $\Psi(T)$ of a forest $T$ forms a greedoid on its vertex set. The
cases where $G$ is bipartite, triangle-free, well-covered, while $\Psi(G)$ is
a greedoid, were analyzed in \cite{LevMan45}, \cite{LevMan07}, \cite{LevMan08}%
, respectively.

In this paper we demonstrate that if $G$ is a very well-covered graph, then
the family $\Psi(G)$ is a greedoid if and only if $G$ has a unique perfect matching.


\end{abstract}

\section{Introduction}

Throughout this paper $G=(V,E)$ is a simple (i.e., a finite, undirected,
loopless and without multiple edges) graph with vertex set $V=V(G)$ and edge
set $E=E(G)$. If $X\subset V$, then $G[X]$ is the subgraph of $G$ spanned by
$X$. If $A,B$ $\subset V$ and $A\cap B=\emptyset$, then $(A,B)$ stands for the
set $\{e=ab:a\in A,b\in B,e\in E\}$. The \textit{neighborhood} of a vertex
$v\in V$ is the set $N(v)=\{u:u\in V$\ \textit{and} $vu\in E\}$. For $A\subset
V$, we denote $N_{G}(A)=\{v\in V-A:N(v)\cap A\neq\emptyset\}$ and
$N_{G}[A]=A\cup N(A)$, or shortly, $N(A)$ and $N[A]$. If $N(v)=\{u\}$, then
$v$ is a \textit{pendant vertex} and $uv$ is a \textit{pendant edge} of $G$.

$K_{n},C_{n},P_{n}$ denote respectively, the complete graph on $n\geq1$
vertices, the chordless cycle on $n\geq3$ vertices, and the chordless path on
$n\geq2$ vertices.

A \textit{matching} in a graph $G=(V,E)$ is a set $M\subseteq E$ such that no
two edges of $M$ share a common vertex. A \textit{maximum matching} is a
matching of maximum cardinality. By $\mu(G)$ is denoted the size of a maximum
matching. A matching is \textit{perfect} if it saturates all the vertices of
the graph.

\begin{figure}[h]
\setlength{\unitlength}{1.0cm} \begin{picture}(5,1.5)\thicklines
\multiput(3,0)(1,0){4}{\circle*{0.29}}
\multiput(4,1)(1,0){3}{\circle*{0.29}}
\put(3,0){\line(1,0){3}}
\put(5,0){\line(1,1){1}}
\put(4,1){\line(1,0){1}}
\multiput(4,0)(1,0){2}{\line(0,1){1}}
\put(2.7,0){\makebox(0,0){$a$}}
\put(3.7,0.3){\makebox(0,0){$b$}}
\put(4.7,0.3){\makebox(0,0){$c$}}
\put(6.3,0){\makebox(0,0){$d$}}
\put(3.7,1){\makebox(0,0){$e$}}
\put(5.3,1){\makebox(0,0){$f$}}
\put(6.3,1){\makebox(0,0){$g$}}
\put(2.1,0.5){\makebox(0,0){$G$}}
\multiput(8.5,0)(1,0){3}{\circle*{0.29}}
\multiput(9.5,1)(1,0){3}{\circle*{0.29}}
\put(8.5,0){\line(1,0){2}}
\put(9.5,1){\line(1,0){2}}
\multiput(9.5,0)(1,0){2}{\line(0,1){1}}
\put(8.2,0){\makebox(0,0){$u$}}
\put(9.2,0.3){\makebox(0,0){$v$}}
\put(10.8,0){\makebox(0,0){$t$}}
\put(11.85,1){\makebox(0,0){$w$}}
\put(9.2,1){\makebox(0,0){$y$}}
\put(10.8,0.7){\makebox(0,0){$x$}}
\put(7.7,0.5){\makebox(0,0){$H$}}
\end{picture}\caption{The unique cycle of $H$ is alternating with respect to
the matching $\{yv,tx\}$.}%
\label{fig111}%
\end{figure}
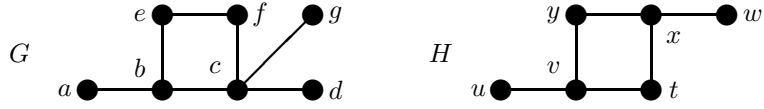

If for every two incident edges of a cycle $C$ exactly one of them belongs to
a matching $M$, then $C$ is called an $M$\textit{-alternating cycle
}\cite{Krogdahl}. It is clear that an $M$-alternating cycle should be of even
length. A matching $M$ in $G$ is called \textit{alternating cycle-free} if $G$
has no $M$-alternating cycle. Alternating cycle-free matchings for bipartite
graphs were first defined in \cite{Krogdahl}. For example, the matching
$\{ab,cd,ef\}$ of the graph $G$ from Figure \ref{fig111} is alternating
cycle-free.

A matching $M=\{a_{i}b_{i}:a_{i},b_{i}\in V(G),1\leq i\leq k\}$ of graph $G$
is called \textit{a uniquely restricted matching} if $M$ is the unique perfect
matching of $G[\{a_{i},b_{i}:1\leq i\leq k\}]$ \cite{GolHiLew}.

\begin{theorem}
\cite{GolHiLew}\label{th9} A matching $M$ in a graph $G$ is uniquely
restricted if and only if $G$ contains no alternating cycle with respect to
$M$, i.e., $M$ is \textit{alternating} cycle-free.
\end{theorem}

For instance, all the maximum matchings of the graph $G$ in Figure
\ref{fig111} are uniquely restricted, while the graph $H$ from the same figure
has both uniquely restricted maximum matchings (e.g., $\{uv,xw\}$) and
non-uniquely restricted maximum matchings (e.g., $\{xy,tv\}$).

A \textit{stable} set in $G$ is a set of pairwise non-adjacent vertices. A
stable set of maximum size will be referred to as a \textit{maximum stable
set} of $G$, and the \textit{stability number }of $G$, denoted by $\alpha(G)$,
is the cardinality of a maximum stable set in $G$. Let $\Omega(G)$ stand for
the set of all maximum stable sets of $G$.

In general, $\alpha(G)\leq\alpha(G-e)$ and $\mu(G-e)\leq\mu(G)$ holds for any
edge $e$ of a graph $G$. An edge $e$ of $G$ is $\alpha$\textit{-critical}
($\mu$\textit{-critical}) if $\alpha(G)<\alpha(G-e)$ ($\mu(G)>\mu(G-e)$,
respectively). It is worth observing that there is no general connection
between the $\alpha$-critical and $\mu$-critical edges of a graph.

Recall that $G$ is called a \textit{K\"{o}nig-Egerv\'{a}ry graph} provided
$\alpha(G)+\mu(G)=\left\vert V(G)\right\vert $ \cite{Dem}, \cite{Ster}. As a
well-known example, every bipartite graph is a K\"{o}nig-Egerv\'{a}ry graph
\cite{Eger}, \cite{Koen}.

\begin{theorem}
\label{th4}If $G$ is a \textit{K\"{o}nig-Egerv\'{a}ry graph, then the
following assertions hold:}

\emph{(i)} \cite{LevMan2003} \textit{every maximum matching is contained in
}$\left(  S,V(G)-S\right)  $, for each $S\in\Omega\left(  G\right)  $;

\emph{(ii)} \cite{LevMan06} the $\alpha$-critical edges are also $\mu
$-critical, and they coincide in a bipartite graph.
\end{theorem}

A set $A\subseteq V(G)$ is a \textit{local maximum stable set} of $G$ if
$A\in\Omega(G[N[A]])$ \cite{LevMan2}; by $\Psi(G)$ we denote the set of all
local maximum stable sets of the graph $G$.

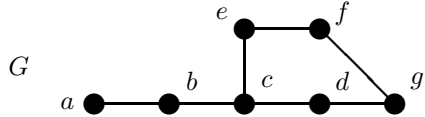
\begin{figure}[h]
\setlength{\unitlength}{1.0cm} \begin{picture}(5,1.5)\thicklines
\multiput(4.5,0)(1,0){5}{\circle*{0.29}}
\multiput(6.5,1)(1,0){2}{\circle*{0.29}}
\put(4.5,0){\line(1,0){4}}
\put(6.5,1){\line(1,0){1}}
\put(6.5,0){\line(0,1){1}}
\put(7.5,1){\line(1,-1){1}}
\put(4.15,0){\makebox(0,0){$a$}}
\put(5.8,0.3){\makebox(0,0){$b$}}
\put(6.8,0.3){\makebox(0,0){$c$}}
\put(7.8,0.3){\makebox(0,0){$d$}}
\put(6.2,1.2){\makebox(0,0){$e$}}
\put(7.8,1.2){\makebox(0,0){$f$}}
\put(8.8,0.3){\makebox(0,0){$g$}}
\put(3.5,0.5){\makebox(0,0){$G$}}
\end{picture}\caption{$\{a\},\{e,d\},\{a,d,f\}\in\Psi\left(  G\right)  $,
while $\{b\},\{a,e\},\{c,f\}$ are not in $\Psi\left(  G\right)  $.}%
\label{fig10}%
\end{figure}

The following theorem concerning maximum stable sets in general graphs, due to
Nemhauser and Trotter Jr. \cite{NemhTro}, shows that for a special subgraph
$H$ of a graph $G$, some maximum stable set of $H$ can be enlarged to a
maximum stable set of $G$.

\begin{theorem}
\cite{NemhTro}\label{th1} Every local maximum stable set of a graph is a
subset of a maximum stable set.
\end{theorem}

Let us notice that the converse of Theorem \ref{th1} is not generally true.
For instance, $C_{n}$ has no proper local maximum stable set, for any $n\geq
4$. The graph $G$ in Figure \ref{fig10} shows another counterexample: any
$S\in\Omega(G)$ contains some local maximum stable set, but these local
maximum stable sets are of different cardinalities. As examples,
$\{a,d,f\}\in\Omega(G)$ and $\{a\},\{d,f\}\in\Psi(G)$, while for
$\{b,e,g\}\in\Omega(G)$ only $\{e,g\}\in\Psi(G)$.

\begin{definition}
\cite{BjZiegler}, \cite{KorLovSch} A \textit{greedoid} is a pair
$(V,\mathcal{F})$, where $\mathcal{F}\subseteq2^{V}$ is a non-empty set system
satisfying the following conditions:

\emph{Accessibility:} for every non-empty $X\in\mathcal{F}$ there is an $x\in
X$ such that $X-\{x\}\in\mathcal{F}$;

\emph{Exchange:} for $X,Y\in\mathcal{F},\left\vert X\right\vert =\left\vert
Y\right\vert +1$, there is an $x\in X-Y$ such that $Y\cup\{x\}\in\mathcal{F}$.
\end{definition}

\begin{definition}
\cite{LevMan08b} A greedoid $(V,\mathcal{F})$ is called a \textit{local
maximum stable set greedoid} if there exists a graph $G=(V,E)$ such that
$\mathcal{F}=\Psi(G)$.
\end{definition}

In fact, the following theorem says that, in the case of local maximum stable
set greedoids, it is enough to check only the accessibility property.

\begin{theorem}
\label{th7}\cite{LevMan08b} If the family $\Psi(G)$ of a graph $G$ satisfies
the accessibility property, then $\left(  V(G),\Psi(G)\right)  $ is a greedoid.
\end{theorem}

In the sequel, we use $\mathcal{F}$ instead of $(V,\mathcal{F})$, as the
ground set $V$ will be, usually, the vertex set of some graph.

\begin{theorem}
\label{th2}\cite{LevMan2} The family of local maximum stable sets of a forest
forms a greedoid on its vertex set.
\end{theorem}

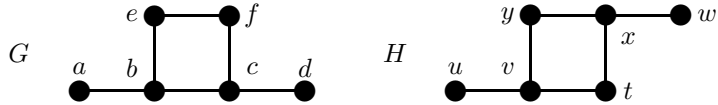
\begin{figure}[h]
\setlength{\unitlength}{1.0cm} \begin{picture}(5,1.5)\thicklines
\multiput(3,0)(1,0){4}{\circle*{0.29}}
\multiput(4,1)(1,0){2}{\circle*{0.29}}
\put(3,0){\line(1,0){3}}
\put(4,1){\line(1,0){1}}
\multiput(4,0)(1,0){2}{\line(0,1){1}}
\put(3,0.3){\makebox(0,0){$a$}}
\put(3.7,0.3){\makebox(0,0){$b$}}
\put(5.3,0.3){\makebox(0,0){$c$}}
\put(6,0.3){\makebox(0,0){$d$}}
\put(3.7,1){\makebox(0,0){$e$}}
\put(5.3,1){\makebox(0,0){$f$}}
\put(2.2,0.5){\makebox(0,0){$G$}}
\multiput(8,0)(1,0){3}{\circle*{0.29}}
\multiput(9,1)(1,0){3}{\circle*{0.29}}
\put(8,0){\line(1,0){2}}
\put(9,1){\line(1,0){2}}
\multiput(9,0)(1,0){2}{\line(0,1){1}}
\put(8,0.3){\makebox(0,0){$u$}}
\put(8.7,0.3){\makebox(0,0){$v$}}
\put(10.3,0){\makebox(0,0){$t$}}
\put(11.35,1){\makebox(0,0){$w$}}
\put(8.7,1){\makebox(0,0){$y$}}
\put(10.3,0.7){\makebox(0,0){$x$}}
\put(7.2,0.5){\makebox(0,0){$H$}}
\end{picture}
\caption{Both $G$ and $H$ are bipartite, but only $\Psi(G)$ forms {a
greedoid}.}%
\label{Fig101}%
\end{figure}

The conclusion of Theorem \ref{th2} is not specific for forests. For instance,
the family $\Psi(G)$ of the graph $G$ in Figure \ref{Fig101} is a greedoid.

Notice that $\Psi(H)$ is not a greedoid, where $H$ is from Figure
\ref{Fig101}, because the accessibility property is not satisfied; e.g.,
$\{y,t\}\in\Psi(H)$, while $\{y\},\{t\}$ $\notin\Psi(H)$. In addition, one can
see that all the maximum matchings of the graph $G$ in Figure \ref{Fig101} are
uniquely restricted, while the graph $H$ from the same figure has both
uniquely restricted maximum matchings (e.g., $\{uv,xw\}$) and non-uniquely
restricted maximum matchings (e.g., $\{xy,tv\}$). It turns out that this is
the reason that $\Psi(H)$ is not a greedoid, while $\Psi(G)$ is a greedoid.

\begin{theorem}
\label{th22}\cite{LevMan45} For a bipartite graph $G,$ $\Psi(G)$ is a greedoid
on its vertex set if and only if all its maximum matchings are uniquely restricted.
\end{theorem}

The case of bipartite graphs owning a unique cycle, whose family of local
maximum stable sets forms a greedoid is analyzed in \cite{LevMan5}.

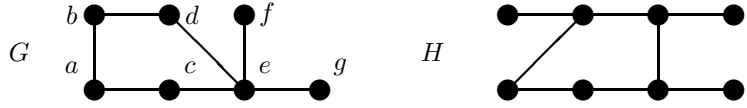
\begin{figure}[h]
\setlength{\unitlength}{1.0cm} \begin{picture}(5,1.3)\thicklines
\multiput(3,0)(1,0){4}{\circle*{0.29}}
\multiput(3,1)(1,0){3}{\circle*{0.29}}
\put(3,0){\line(1,0){3}}
\put(3,1){\line(1,0){1}}
\put(4,1){\line(1,-1){1}}
\put(3,0){\line(0,1){1}}
\put(5,0){\line(0,1){1}}
\put(5.3,1){\makebox(0,0){$f$}}
\put(2.7,0.3){\makebox(0,0){$a$}}
\put(2.7,1){\makebox(0,0){$b$}}
\put(4.27,0.3){\makebox(0,0){$c$}}
\put(4.3,1){\makebox(0,0){$d$}}
\put(5.27,0.3){\makebox(0,0){$e$}}
\put(6.27,0.3){\makebox(0,0){$g$}}
\put(2,0.5){\makebox(0,0){$G$}}
\multiput(8.5,0)(1,0){4}{\circle*{0.29}}
\multiput(8.5,1)(1,0){4}{\circle*{0.29}}
\put(8.5,0){\line(1,0){3}}
\put(8.5,0){\line(1,1){1}}
\put(8.5,1){\line(1,0){3}}
\put(10.5,0){\line(0,1){1}}
\put(7.5,0.5){\makebox(0,0){$H$}}
\end{picture}\caption{$\Psi(G)$ is not a greedoid, $\Psi(H)$ is a greedoid.}%
\label{fig2922}%
\end{figure}

The graphs from Figure \ref{fig2922} are non-bipartite K\"{o}nig-Egerv\'{a}ry
graphs, and all their maximum matchings are uniquely restricted. Let us remark
that both graphs are also triangle-free, but only $\Psi(H)$ is a greedoid. It
is clear that $\{b,c\}\in$ $\Psi(G)$, while $G[N[\{b,c\}]]$ is not a
K\"{o}nig-Egerv\'{a}ry graph. As one can see from the following theorem, this
observation is the real reason for $\Psi(G)$ not to be a greedoid.

\begin{theorem}
\label{th33}\cite{LevMan07} If $G$ is a triangle-free graph, then $\Psi(G)$ is
a greedoid if and only if all maximum matchings of $G$ are uniquely restricted
and the closed neighborhood of every local maximum stable set of $G$ induces a
K\"{o}nig-Egerv\'{a}ry graph.
\end{theorem}

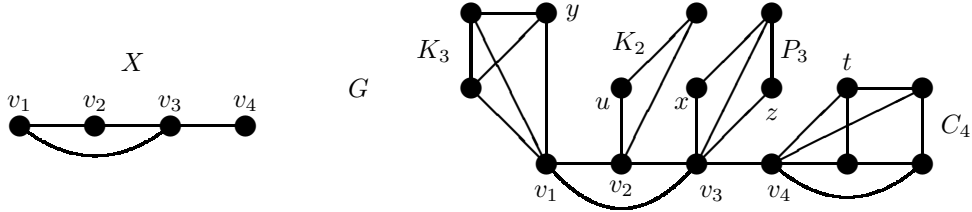
\begin{figure}[h]
\setlength{\unitlength}{1cm}\begin{picture}(5,2.8)\thicklines
\multiput(1,1)(1,0){4}{\circle*{0.29}}
\put(1,1){\line(1,0){3}}
\qbezier(1,1)(2,0.2)(3,1)
\put(1,1.3){\makebox(0,0){$v_1$}}
\put(2,1.3){\makebox(0,0){$v_2$}}
\put(3,1.3){\makebox(0,0){$v_3$}}
\put(4,1.3){\makebox(0,0){$v_4$}}
\put(2.5,1.85){\makebox(0,0){$X$}}
\multiput(8,0.5)(1,0){6}{\circle*{0.29}}
\put(8,2.5){\circle*{0.29}}
\multiput(7,1.5)(0,1){2}{\circle*{0.29}}
\multiput(9,1.5)(1,0){5}{\circle*{0.29}}
\multiput(10,2.5)(1,0){2}{\circle*{0.29}}
\multiput(8,0.5)(1,0){5}{\line(1,0){1}}
\multiput(9,0.5)(1,0){2}{\line(0,1){1}}
\multiput(9,0.5)(1,0){2}{\line(1,2){1}}
\multiput(9,1.5)(1,0){2}{\line(1,1){1}}
\multiput(10,0.5)(1,0){2}{\line(1,1){1}}
\put(7,1.5){\line(1,1){1}}
\put(7,1.5){\line(0,1){1}}
\put(7,2.5){\line(1,0){1}}
\put(7,1.5){\line(1,-1){1}}
\put(7,2.5){\line(1,-2){1}}
\put(8,0.5){\line(0,1){2}}
\put(11,1.5){\line(0,1){1}}
\put(12,0.5){\line(0,1){1}}
\put(12,1.5){\line(1,0){1}}
\put(13,0.5){\line(0,1){1}}
\put(11,0.5){\line(2,1){2}}
\qbezier(11,0.5)(12,-0.4)(13,0.5)
\qbezier(8,0.5)(9,-0.7)(10,0.5)
\put(9.8,1.3){\makebox(0,0){$x$}}
\put(8.35,2.5){\makebox(0,0){$y$}}
\put(11,1.15){\makebox(0,0){$z$}}
\put(8.75,1.3){\makebox(0,0){$u$}}
\put(12,1.85){\makebox(0,0){$t$}}
\put(8,0.1){\makebox(0,0){$v_1$}}
\put(9,0.17){\makebox(0,0){$v_2$}}
\put(10.2,0.1){\makebox(0,0){$v_3$}}
\put(11.1,0.1){\makebox(0,0){$v_4$}}
\put(6.5,2){\makebox(0,0){$K_3$}}
\put(9.1,2.1){\makebox(0,0){$K_2$}}
\put(11.3,2){\makebox(0,0){$P_3$}}
\put(13.45,1){\makebox(0,0){$C_4$}}
\put(5.5,1.5){\makebox(0,0){$G$}}
\end{picture}\caption{The corona $G=X\circ\{K_{3},K_{2},P_{3},C_{4}\}$.}%
\label{fig121}%
\end{figure}

Let $X$ be a graph with $V(X)=\{v_{i}:1\leq i\leq n\}$, and $\{H_{i}:1\leq
i\leq n\}$ be a family of graphs. Joining each $v_{i}\in V(X)$ to all the
vertices of $H_{i}$, we obtain a new graph, called the \textit{corona} of $X$
and $\{H_{i}:1\leq i\leq n\}$ and denoted by $G=X\circ\{H_{1},H_{2}%
,...,H_{n}\}$. For instance, see Figure \ref{fig121}. If $H_{1}=H_{2}%
=...=H_{n}=H$, we write $G=X\circ H$, and in this case, $G$ is called the
\textit{corona} of $X$ and $H$.

If each $H_{i}$ is a complete graph, then $X\circ\{H_{1},H_{2},...,H_{n}\}$ is
called the \textit{clique corona }of $X$ and $\{H_{1},H_{2},...,H_{n}\}$;
notice that the clique corona is well-covered (and very well-covered, whenever
$H_{i}=K_{1},1\leq i\leq n$). Recall that $G$ is \textit{well-covered} if all
its maximal stable sets have the same cardinality \cite{Plummer}, and $G$ is
\textit{very well-covered} if, in addition, it has no isolated vertices and
$\left\vert V(G)\right\vert =2\alpha(G)$ \cite{Favaron}.

A number of classes of well-covered graphs were completely described (see, for
instance, the following references: \cite{Favaron}, \cite{FinHartNow},
\cite{HartPlummer}, \cite{LevMan98}, \cite{PrisToppVest}, \cite{Ravindra},
\cite{Stap}.

\begin{theorem}
\label{th88}\emph{(i)} \cite{FinHartNow} Let $G$ be a connected graph of girth
$\geq6$, which is isomorphic to neither $C_{7}$ nor $K_{1}$. Then $G$ is
well-covered if and only if $G=H\circ K_{1}$, for some graph $H$ of girth
$\geq6$.

\emph{(ii)} \cite{DeanZio1994}, \cite{LevMan07a} Let $G$ be a graph having
girth $\geq5$. Then $G$ is very well-covered if and only if $G=H\circ K_{1}$,
for some graph $H$ of girth $\geq5$.

\emph{(iii) }\cite{LevMan98} $G$ is very well-covered if and only if $G$ is a
well-covered K\"{o}nig-Egerv\'{a}ry graph.

\emph{(iv)} \cite{ToppVolk} $G=X\circ\{H_{1},H_{2},...,H_{n}\}$ is
well-covered if and only if all $H_{i}$ are complete.
\end{theorem}

\begin{figure}[h]
\setlength{\unitlength}{1cm}\begin{picture}(5,1.4)\thicklines
\multiput(4,0)(1,0){2}{\circle*{0.29}}
\multiput(4,1)(1,0){2}{\circle*{0.29}}
\put(4,0){\line(1,0){1}}
\put(4,1){\line(1,0){1}}
\put(4,0){\line(0,1){1}}
\put(5,0){\line(0,1){1}}
\put(3,0.5){\makebox(0,0){$C_4$}}
\multiput(7.5,0)(1,0){4}{\circle*{0.29}}
\multiput(8.5,1)(1,0){2}{\circle*{0.29}}
\multiput(8.5,0)(1,0){2}{\line(0,1){1}}
\put(7.5,0){\line(1,0){3}}
\put(8.5,1){\line(1,0){1}}
\put(6.5,0.5){\makebox(0,0){$G$}}
\end{picture}\caption{Very well-covered graphs with no perfect mathching
consisting of only pendant edges.}%
\label{fig81}%
\end{figure}
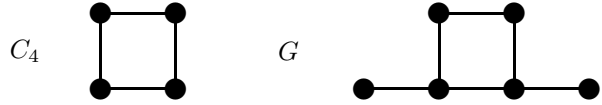

It is easy to prove that every graph having a perfect matching consisting of
pendant edges is very well-covered. The converse is not generally true (see,
for instance, the graphs depicted in Figure \ref{fig81}). Moreover, there are
well-covered graphs without perfect matchings; e.g., $K_{3}$.

\begin{theorem}
\label{th11}\cite{Favaron}\ For a graph $G$ without isolated vertices the
following are equivalent:

\emph{(i)} $G$ is very well-covered;

\emph{(ii)} there exists a perfect matching $M$ in $G$ that satisfies property
$P$, i.e.,
\[
\text{\textquotedblleft}N(x)\cap N(y)=\emptyset\text{, and each }v\in
N(x)-\{y\}\text{ is adjacent to all vertices of }N(y)-\left\{  x\right\}
\text{\textquotedblright}%
\]
hold for every edge $xy\in M$;

\emph{(iii)} there exists at least one perfect matching in $G$ and every
perfect matching in $G$ satisfies property $P$.
\end{theorem}

Various cases of well-covered graphs generating local maximum stable set
greedoids, were treated in \cite{LevMan08}, \cite{LevMan08a}, \cite{LevMan09},
\cite{LevMan10}.

\begin{theorem}
\label{th10}Let $G=X\circ\{H_{1},H_{2},...,H_{n}\}$, where $H_{1}%
,H_{2},...,H_{n}$ are non-empty graphs.

\emph{(i)} \cite{LevMan09} if $G=P_{n}$ and all $H_{i},1\leq i\leq n$, are
complete graphs, then $\Psi(G)$ is a greedoid;

\emph{(ii)} \cite{LevMan08} if $H_{i}=K_{1},1\leq i\leq n$, then $\Psi(G)$ is
a greedoid;

\emph{(iii)} \cite{LevMan08a} if all $H_{1},H_{2},...,H_{n}$ are complete
graphs, then $\Psi(G)$ is a greedoid;

\emph{(iv)} \cite{LevMan10} $\Psi(G)$ is a greedoid if and only if every
$\Psi(H_{i}),i=1,2,...,n$, is a greedoid.
\end{theorem}

It turns out that the property of having a unique maximum matching is of
crucial importance for very-well covered graphs to generate local maximum
stable set greedoids.

\begin{theorem}
\cite{LevMan10a} Let $G$ be a very well-covered graph of girth at least $4$.
Then $\Psi(G)$ is a greedoid if and only if $G$ has a unique maximum matching.
\end{theorem}

In this paper we completely characterize very well-covered graphs whose
families of local maximum stable sets are greedoids.

\section{Very well-covered graphs producing greedoids}

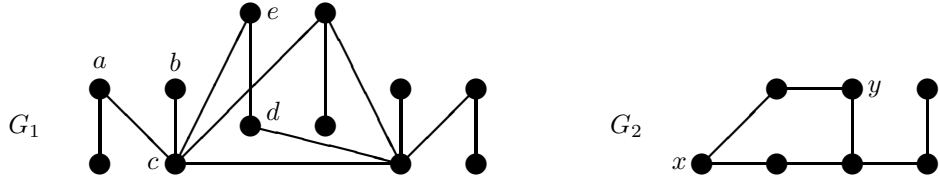
\begin{figure}[h]
\setlength{\unitlength}{1cm}\begin{picture}(5,2.5)\thicklines
\multiput(2,0)(1,0){2}{\circle*{0.29}}
\multiput(2,1)(1,0){2}{\circle*{0.29}}
\multiput(6,0)(1,0){2}{\circle*{0.29}}
\multiput(6,1)(1,0){2}{\circle*{0.29}}
\multiput(4,2)(1,0){2}{\circle*{0.29}}
\multiput(4,0.5)(1,0){2}{\circle*{0.29}}
\put(2,0){\line(0,1){1}}
\put(2,1){\line(1,-1){1}}
\put(3,0){\line(0,1){1}}
\put(3,0){\line(1,0){3}}
\put(3,0){\line(1,2){1}}
\put(3,0){\line(1,1){2}}
\put(4,2){\line(0,-1){1.5}}
\put(4,0.5){\line(4,-1){2}}
\put(5,2){\line(0,-1){1.5}}
\put(5,2){\line(1,-2){1}}
\put(6,0){\line(0,1){1}}
\put(6,0){\line(1,1){1}}
\put(7,0){\line(0,1){1}}
\put(2,1.35){\makebox(0,0){$a$}}
\put(3,1.35){\makebox(0,0){$b$}}
\put(2.7,0){\makebox(0,0){$c$}}
\put(4.3,0.7){\makebox(0,0){$d$}}
\put(4.3,2){\makebox(0,0){$e$}}
\put(1,0.5){\makebox(0,0){$G_{1}$}}
\multiput(10,0)(1,0){4}{\circle*{0.29}}
\multiput(11,1)(1,0){3}{\circle*{0.29}}
\put(10,0){\line(1,0){3}}
\put(10,0){\line(1,1){1}}
\put(11,1){\line(1,0){1}}
\put(12,0){\line(0,1){1}}
\put(13,0){\line(0,1){1}}
\put(9.7,0){\makebox(0,0){$x$}}
\put(12.3,1){\makebox(0,0){$y$}}
\put(9,0.5){\makebox(0,0){$G_{2}$}}
\end{picture}\caption{Non-bipartite well-covered graphs. Moreover, $G_{1}$ is
very well-covered.}%
\label{fig31}%
\end{figure}

Notice that $S_{1}=\{a,b\}$ and $S_{2}=\{c,d\}$ are stable sets in the graph
$G_{1}$ from Figure \ref{fig31}, $S_{1}\in\Psi(G_{1})$, and both
$G_{1}[N[S_{1}]]$ and $G_{1}[N[S_{2}]]$ are K\"{o}nig-Egerv\'{a}ry graphs. On
the other hand, $S_{2}=\{x,y\}\in\Psi(G_{2})$, where $G_{2}$ is from Figure
\ref{fig31}, but $G[N[S_{3}]]$ is not a K\"{o}nig-Egerv\'{a}ry graph.

\begin{theorem}
\cite{LevMan07a}\label{th3} Let $G$ be a very well-covered graph. Then
$G[N[S]]$ is a K\"{o}nig-Egerv\'{a}ry graph, for every $S\in\Psi(G)$.
\end{theorem}

Concerning the graph $G_{1}$ from Figure \ref{fig31}, let us remark that
$\{b,d\},\{b,e\}$ are stable sets, $\left\vert \{b,d\}\right\vert <\left\vert
N(\{b,d\})\right\vert $ and $\left\vert \{b,e\}\right\vert =\left\vert
N(\{b,e\})\right\vert $, but only $\{b,e\}\in\Psi(G_{1})$.

\begin{lemma}
\label{lem3}If $S$ is a stable set in a very well-covered graph $G$, then
$S\in\Psi(G)$ if and only if $\left\vert S\right\vert =\left\vert
N(S)\right\vert $.
\end{lemma}

\begin{proof}
According to Theorems \ref{th88}\emph{(iii)} and \ref{th11}, $G$ is a
K\"{o}nig-Egerv\'{a}ry graph having a perfect matching, say $M$. If $S$ is a
stable set in $G$, there must be some $A\in\Omega(G)$, such that $S\subseteq
A$, because $G$ is well-covered. By Theorem \ref{th4}\emph{(i)}, we have that
$M\subseteq(A,V(G)-A)$. Since $M$ is a perfect matching, it follows that $S$
is matched into $N(S)$, and further,
\[
\left\vert S\right\vert =\left\vert M(S)\right\vert \leq\left\vert
N(S)\right\vert ,
\]
where $M(S)=\{y\in V:xy\in M,x\in S\}$.

Let $S\in\Psi(G)$. According to Theorem \ref{th3}, $G[N[S]]$ is a
K\"{o}nig-Egerv\'{a}ry graph, and consequently, we get that $\left\vert
S\right\vert =\left\vert M(S)\right\vert \geq\left\vert N(S)\right\vert $.
Hence, we infer that $\left\vert S\right\vert =\left\vert N(S)\right\vert $.

Conversely, let $S$ be a stable set in $G$ satisfying $\left\vert S\right\vert
=\left\vert N(S)\right\vert $. Since $S$ is matched by $M$ into $N(S)$, we
infer that the restriction of $M$ to $G[N[S]]$ is a perfect matching.
Therefore, $\left\vert S\right\vert =\alpha(G[N[S]])$, and this implies
$S\in\Psi(G)$.
\end{proof}

\begin{figure}[h]
\setlength{\unitlength}{1cm}\begin{picture}(5,1.2)\thicklines
\multiput(2,0)(1,0){5}{\circle*{0.29}}
\multiput(5,1)(1,0){2}{\circle*{0.29}}
\put(3,1){\circle*{0.29}}
\put(2,0){\line(1,1){1}}
\put(3,0){\line(0,1){1}}
\put(2,0){\line(1,0){4}}
\put(4,0){\line(1,1){1}}
\put(5,1){\line(1,0){1}}
\put(6,0){\line(0,1){1}}
\put(5,0.3){\makebox(0,0){$x$}}
\put(6.3,1){\makebox(0,0){$y$}}
\put(3.3,1){\makebox(0,0){$v$}}
\put(1.2,0.5){\makebox(0,0){$G_{1}$}}
\multiput(8,0)(1,0){5}{\circle*{0.29}}
\multiput(9,1)(1,0){2}{\circle*{0.29}}
\multiput(12,1)(1,0){2}{\circle*{0.29}}
\put(8,0){\line(1,0){4}}
\put(9,0){\line(0,1){1}}
\put(9,1){\line(1,0){1}}
\put(10,1){\line(1,-1){1}}
\put(12,0){\line(0,1){1}}
\put(12,0){\line(1,1){1}}
\put(7.7,0){\makebox(0,0){$a$}}
\put(8.7,0.3){\makebox(0,0){$b$}}
\put(8.7,1){\makebox(0,0){$c$}}
\put(10.3,1){\makebox(0,0){$d$}}
\put(10,0.3){\makebox(0,0){$e$}}
\put(11,0.35){\makebox(0,0){$f$}}
\put(12.3,0){\makebox(0,0){$u$}}
\put(11.7,1){\makebox(0,0){$v$}}
\put(13.3,1){\makebox(0,0){$w$}}
\put(7.2,0.5){\makebox(0,0){$G_{2}$}}
\end{picture}\caption{$G_{1}$ and $G_{2}$ are not very well-covered graphs.
Only $G_{1}$ is well-covered.}%
\label{fig121212}%
\end{figure}
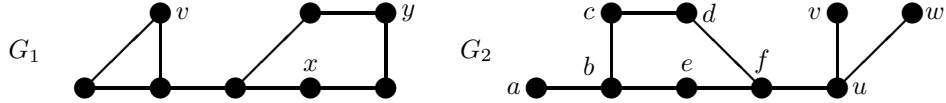Notice that the above lemma can fail in a non-very well-covered
graph. For instance, $S=\{x,y\}\in\Psi(G_{1})$, while $\left\vert S\right\vert
<\left\vert N(S)\right\vert $, where $G_{1}$ is from Figure \ref{fig121212}
and it is well-covered. Further, the sets $S_{1}=\{a,c\},S_{2}=\{e,d\}$ and
$S_{3}=\{v,w\}$ belong to $\Psi(G_{2})$, where $G_{2}$ is from Figure
\ref{fig121212}, and they satisfy:
\[
\left\vert S_{1}\right\vert =\left\vert N(S_{1})\right\vert ,\left\vert
S_{2}\right\vert <\left\vert N(S_{2})\right\vert ,\text{ and }\left\vert
S_{3}\right\vert >\left\vert N(S_{3})\right\vert .
\]

Concerning the very well-covered graph $G_{1}$ from Figure \ref{fig31}, we see
that $A,B\in\Psi(G_{1})$, where $B=\{a\},A=B\cup\{a\}$, and $\left\vert
N(A)\right\vert =\left\vert N(B)\right\vert +1$. The following lemma shows
that in a very well-covered graph the existence of an accessibility chain is
equivalent to the fact that one can have a chain of stable sets, where each
additional vertex added to a stable set increases the size of its open
neighborhood by exactly one element.

\begin{lemma}
\label{lem65}If $A=B\cup\{v\}$ is a stable set in a very well-covered graph
$G$, and $B\in\Psi(G)$, then $A\in\Psi(G)$ if and only if $\left\vert
N(A)\right\vert =\left\vert N(B)\right\vert +1$.
\end{lemma}

\begin{proof}
Assume that $A=B\cup\{v\}\in\Psi(G)$. Since $G$ is very well-covered, by Lemma
\ref{lem3}, it follows that $\left\vert N(A)\right\vert -\left\vert
N(B)\right\vert =\left\vert A\right\vert -\left\vert B\right\vert =1$.

Conversely, since
\[
\left\vert N(A)\right\vert =\left\vert N(B)\right\vert +1=\left\vert
B\right\vert +1=\left\vert A\right\vert ,
\]
Lemma \ref{lem3} implies immediately that $A\in\Psi(G)$.
\end{proof}

Lemma \ref{lem65} fails for graphs that are not very well-covered, for
instance, $B_{1}=\{x,y\}$ and $A_{1}=B_{1}\cup\{v\}$ belong to $\Psi(G_{1})$,
but $\left\vert N(A_{1})\right\vert =\left\vert N(B_{1})\right\vert +2$, where
$G_{1}$ is from Figure \ref{fig121212}, and also $B_{2}=\{v\}$ and
$A_{2}=B_{2}\cup\{w\}$ belong to $\Psi(G_{2})$, but $\left\vert N(A_{2}%
)\right\vert =\left\vert N(B_{2})\right\vert $, where $G_{2}$ is from Figure
\ref{fig121212}.

\begin{figure}[h]
\setlength{\unitlength}{1cm}\begin{picture}(5,1.2)\thicklines
\multiput(3.5,0)(1,0){3}{\circle*{0.29}}
\multiput(3.5,1)(1,0){3}{\circle*{0.29}}
\multiput(1.5,0.5)(1,0){2}{\circle*{0.29}}
\put(1.5,0.5){\line(1,0){1}}
\put(2.5,0.5){\line(4,1){2}}
\put(2.5,0.5){\line(4,-1){2}}
\put(3.5,0){\line(1,0){2}}
\put(3.5,1){\line(1,0){2}}
\put(4.5,0){\line(0,1){1}}
\put(5.5,0){\line(0,1){1}}
\put(0.7,0.5){\makebox(0,0){$G_{1}$}}
\multiput(7.5,0)(1,0){2}{\circle*{0.29}}
\multiput(7.5,1)(1,0){2}{\circle*{0.29}}
\put(7.5,0){\line(1,0){1}}
\put(7.5,0){\line(0,1){1}}
\put(7.5,1){\line(1,0){1}}
\put(8.5,0){\line(0,1){1}}
\put(6.7,0.5){\makebox(0,0){$G_{2}$}}
\multiput(10.5,0)(2,0){2}{\circle*{0.29}}
\multiput(10.5,1)(2,0){2}{\circle*{0.29}}
\put(11.5,0.5){\circle*{0.29}}
\put(13.5,0){\circle*{0.29}}
\put(10.5,0){\line(1,0){3}}
\put(10.5,0){\line(0,1){1}}
\put(10.5,1){\line(1,0){2}}
\put(10.5,0){\line(2,1){2}}
\put(12.5,0){\line(0,1){1}}
\put(9.7,0.5){\makebox(0,0){$G_{3}$}}
\end{picture}\caption{$G_{1},G_{2},G_{3}$ are very well-covered\ graphs.
$G_{1}$ has a unique perfect matching.}%
\label{fig1212}%
\end{figure}
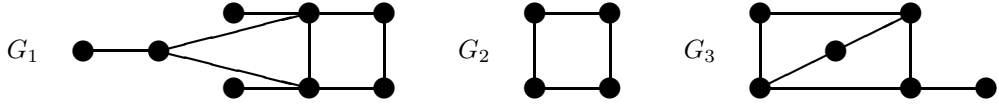

Let us notice that the graphs $G_{1}$,$G_{2}$ and $G_{3}$ from Figure
\ref{fig1212} are very well-covered; by Theorem \ref{th22} or \ref{th33},
neither $\Psi(G_{2})$ nor $\Psi(G_{3})$ is a greedoid. However, $\Psi(G_{1})$
is a greedoid.

\begin{theorem}
\label{th8}Let $G$ be a very well-covered graph. Then $\Psi(G)$ forms a
greedoid if and only if $G$ has a unique maximum matching.
\end{theorem}

\begin{proof}
Suppose that $\Psi(G)$ forms a greedoid. By Theorem \ref{th11},\ $G$ has at
least one perfect matching, say $M$.

Since $\Psi(G)$ is a greedoid, every $S\in\Omega(G)$ has an accessibility
chain
\[
\{x_{1}\}\subset\{x_{1},x_{2}\}\subset...\subset\{x_{1},x_{2},...,x_{\alpha
-1}\}\subset\{x_{1},x_{2},...,x_{\alpha}\}=S.
\]
Let us denote $S_{i}=\{x_{1},x_{2},...,x_{i}\},1\leq i\leq\alpha$, and
$S_{0}=\emptyset$.

Since $S_{i-1}\in\Psi(G),S_{i}=S_{i-1}\cup\{x_{i}\}\in\Psi(G)$ and $G$ is
well-covered, Lemma \ref{lem65} implies that $\left\vert N(x_{i}%
)-N[S_{i-1}]\right\vert =1$, because\textbf{ }%
\[
\left\vert N(x_{i})-N[S_{i-1}]\right\vert =\left\vert N(S_{i})-N(S_{i-1}%
)\right\vert =\left\vert N(S_{i})\right\vert -\left\vert N(S_{i-1})\right\vert
.
\]
Let $\{y_{i}\}=N(x_{i})-N[S_{i-1}],1\leq i\leq\alpha$. Hence, $M=\{x_{i}%
y_{i}:1\leq i\leq\alpha\}$ is a maximum matching in $G$.

Let us validate that $M$ is a uniquely restricted maximum matching in $G$.

We induct on $k=\left\vert S_{k}\right\vert $ in order to show that the
restriction of $M$ to $H_{k}=G[N[S_{k}]]$, which we denote by $M_{k}$, is a
uniquely restricted maximum matching in $H_{k}$.

For $k=1,S_{1}=\{x_{1}\}\in\Psi(G)$ and this implies that $N(x_{1})=\{y_{1}%
\}$, unless $x_{1}$ is an isolated vertex. In this case, $M_{1}=\{x_{1}%
y_{1}\}$ is a uniquely restricted maximum matching in $H_{1}$. If $x_{1}$ is
an isolated vertex, then $M_{1}=$ $\emptyset$ is a uniquely restricted maximum
matching in $H_{1}$.

Suppose that the assertion is true for all $j\leq k-1$. Let us notice that
\[
N[S_{k}]=N[S_{k-1}]\cup\{x_{k}\}\cup\{y_{k}\}.
\]

As we know, $N(x_{k})-N[S_{k-1}]=\{y_{k}\}$.

Since $H_{k}$ is K\"{o}nig-Egerv\'{a}ry graph, $M_{k}$ is a maximum matching
in $H_{k}$. The edge $x_{k}y_{k}$ is $\alpha$-critical in $H_{k}$, because
$\{y_{k}\}=N(x_{k})-N[S_{k-1}]$, and hence, $x_{k}y_{k}$ is also $\mu
$-critical in $H_{k}$, according to Theorem \ref{th4}\emph{(ii)}. Therefore,
any maximum matching of $H_{k}$ contains the edge $x_{k}y_{k}$. Since
$M_{k}=M_{k-1}\cup\{x_{k}y_{k}\}$ and $M_{k-1}$ is a uniquely restricted
maximum matching in $H_{k-1}=H_{k}-\{x_{k},y_{k}\}$, it follows that $M_{k}$
is a uniquely restricted maximum matching in $H_{k}$.

Conversely, assume that $G$ has a unique perfect matching, say $M$.

We show that $\Psi(G)$ satisfies the accessibility property, i.e., for every
non-empty $X\in\Psi(G)$ there is an $x\in X$ such that $X-\{x\}\in\Psi(G)$.

Let $S\in\Psi(G)$. According to Theorem \ref{th1}, there is some $A\in
\Omega(G)$, such that $S\subseteq A$. By Theorem \ref{th4}\emph{(i)},
$M\subseteq(A,V(G)-A)$ and this implies that $S$ is matched into $N(S)$, and
further, $\left\vert S\right\vert =\left\vert M(S)\right\vert \leq\left\vert
N(S)\right\vert $, where $M(S)=\{y\in V:xy\in M,x\in S\}$. According to
Theorem \ref{th3}, $G[N[S]]$ is a K\"{o}nig-Egerv\'{a}ry graph, and this fact
ensures that $\left\vert S\right\vert =\left\vert M(S)\right\vert
\geq\left\vert N(S)\right\vert $. Hence, we infer that $N(S)=M(S)$.

Suppose that $S$ does not satisfies the accessibility property, i.e.,
$S-\{x\}\notin\Psi(G)$ for every $x\in S$. This implies that $N(S-\{x\})=N(S)$%
, for every $x\in S$. Consequently, each vertex in $N(S)$ has at least two
neighbors in $S$.

We show that there is an even cycle $C$ in $G[N[S]]$, such that half of its
edges are in $M$.

Let $x_{1}y_{1}\in M$ and $x_{1}\in S$. Since $\left\vert N(y_{1})\cap
S\right\vert \geq2$, there is a vertex, say $x_{2}$, belonging to
$N(y_{1})\cap S$.

Let $x_{2}y_{2}\in M$; such an edge exists, because $M$ matches $S$ into
$M(S)$. Now, since $\left\vert N(y_{2})\cap S\right\vert \geq2$, there is a
vertex, say $x_{3}$, belonging to $N(y_{2})\cap S$. If $x_{3}=x_{1}$, then the
cycle $C$ spanned by $\{x_{1},y_{1},x_{2},y_{2}\}$ has half of its edges in
$M$. If $x_{3}\neq x_{1}$, then we consider the edge in $M$ that saturates
$x_{3}$, say $x_{3}y_{3}\in M$. Since $G[N[S]]$ is finite, after a number of
steps, we find some vertex in $N(S)$, say $y_{k}$, that is joined by an edge
to some $x_{j}$ for $j<k$. Clearly, the cycle $C$, with
\[
V(C)=\{x_{i},y_{i}:j\leq i\leq k\}
\]
and
\[
E(C)=\{x_{i},y_{i}:j\leq i\leq k\}\cup\{y_{i}x_{i+1}:1\leq i\leq
k-1\}\cup\{x_{j}y_{k}\}
\]
is even and has half of its edges in $M$. Therefore, $M^{\prime}%
=(M-E(C))\cup(E(C)-M)$ is a perfect matching in $G$ and $M\neq M^{\prime}$, in
contradiction with the uniqueness of $M$ in $G$.

Consequently, $\Psi(G)$ satisfies the accessibility property, and, according
to Theorem \ref{th7}, $\Psi(G)$ is a greedoid.
\end{proof}

Let us remark that the very well-covered graph $G_{1}$ in Figure \ref{fig1212}
has a $C_{3}$ and a $C_{4}$; one edge of $C_{4}$ belongs to the unique perfect
matching $M$ of $G_{1}$, but none of the edges of $C_{3}$ is included in $M$.

\begin{lemma}
\label{lem1}No edge of some $C_{q}$, for $q=3$ or $q\geq5$, belongs to a
perfect matching in a very well-covered graph.
\end{lemma}

\begin{proof}
If the graph $G$ is very well-covered, then by Theorem \ref{th11}, $G$ has a
perfect matching, say $M$, and each perfect matching satisfies Property
\emph{P}.

Let $xy\in M$. Then, Property \emph{P} implies that $N(x)\cap N(y)=\emptyset$,
i.e., $xy$ belongs to no $C_{3}$ in $G$. Further, if $v\in N(x)-\{y\}$ and
$u\in N(y)-\{x\}$, Property \emph{P} assures that $vu\in E(G)$, i.e., $xy$
belongs to no $C_{q}$, for $q\geq5$.
\end{proof}

\begin{figure}[h]
\setlength{\unitlength}{1cm}\begin{picture}(5,1.6)\thicklines
\multiput(1.2,0)(1,0){2}{\circle*{0.29}}
\multiput(1.2,1)(1,0){2}{\circle*{0.29}}
\put(1.2,0){\line(1,0){1}}
\put(1.2,0){\line(0,1){1}}
\put(1.2,1){\line(1,0){1}}
\put(2.2,0){\line(0,1){1}}
\put(0.4,0.4){\makebox(0,0){$G_{1}$}}
\multiput(4,0)(2,0){2}{\circle*{0.29}}
\multiput(4,1)(2,0){2}{\circle*{0.29}}
\put(5,0.5){\circle*{0.29}}
\put(7,0){\circle*{0.29}}
\put(4,0){\line(1,0){3}}
\put(4,0){\line(0,1){1}}
\put(4,1){\line(1,0){2}}
\put(4,0){\line(2,1){2}}
\put(6,0){\line(0,1){1}}
\put(3.3,0.4){\makebox(0,0){$G_{2}$}}
\multiput(9,0)(2,0){3}{\circle*{0.29}}
\multiput(9,1)(2,0){3}{\circle*{0.29}}
\put(9,0){\line(1,0){4}}
\put(9,1){\line(1,0){4}}
\put(9,0){\line(0,1){1}}
\put(11,0){\line(0,1){1}}
\put(13,0){\line(0,1){1}}
\qbezier(9,0)(7.2,2.2)(13,1)
\qbezier(9,1)(14.7,2.2)(13,0)
\put(9.2,0.4){\makebox(0,0){$e_{1}$}}
\put(10,0.8){\makebox(0,0){$e_{2}$}}
\put(10.7,0.4){\makebox(0,0){$e_{3}$}}
\put(12,0.2){\makebox(0,0){$e_{4}$}}
\put(12.7,0.4){\makebox(0,0){$e_{5}$}}
\put(8.35,1.1){\makebox(0,0){$e_{6}$}}
\put(7.85,0.4){\makebox(0,0){$G_{3}$}}
\end{picture}\caption{Very well-covered graphs, each having more than one
perfect matching.}%
\label{fig45}%
\end{figure}
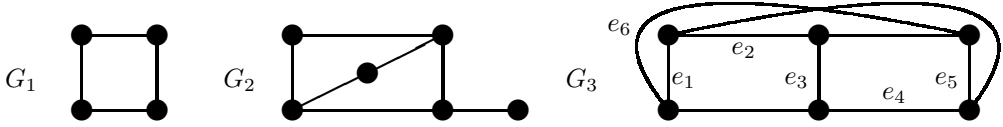

The very well-covered graphs $G_{1}$, $G_{2}$, and $G_{3}$ from Figure
\ref{fig45} have chordless alternating cycles of length $4$. In addition,
$G_{3}$ has an alternating cycle of length $6$, namely, $\{e_{1}%
,e_{2},...,e_{6}\}$ is alternating with respect to the perfect matching
$\{e_{1},e_{3},e_{5}\}$. 

\begin{lemma}
\label{lem2}Let $G$ be a very well-covered graph and $M$ be one of its maximum
matchings. There exists an alternating cycle with respect to $M$ if and only
if there is an alternating chordless cycle of length four with respect to $M$.
\end{lemma}

\begin{proof}
According to Theorem \ref{th11}, every maximum matching of $G$ is perfect.
Suppose $C^{1}$ is an alternating cycle with respect to a perfect matching
$M=\{a_{i}b_{i}:1\leq i\leq\left\vert V(G)\right\vert /2\}$. Without loss of
generality, assume that
\[
C^{1}=\{a_{1},b_{1},a_{2},b_{2},a_{3},...,a_{k-1},b_{k-1},a_{k},b_{k}\}
\]
is a cycle on $2k>4$ vertices with edges%
\[
E(C^{1})=\{a_{1}b_{1},b_{1}a_{2},a_{2}b_{2},b_{2}a_{3},...,a_{k-1}%
b_{k-1},b_{k-1}a_{k},a_{k}b_{k},b_{k}a_{1}\}.
\]

Since $a_{1}b_{1}\in M$, $b_{k}\in N(a_{1})-\{b_{1}\}$, $a_{2}\in
N(b_{1})-\left\{  a_{1}\right\}  $, Property \emph{P} implies that $a_{2}$ is
adjacent to $b_{k}$. Thus the cycle on $2k-2$ vertices $C^{2}=\{a_{2}%
,b_{2},a_{3},...,a_{k-1},b_{k-1},a_{k},b_{k}\}$ with edges
\[
E(C^{2})=\{a_{2}b_{2},b_{2}a_{3},...,a_{k-1}b_{k-1},b_{k-1}a_{k},a_{k}%
b_{k},b_{k}a_{2}\}
\]
is still alternating with respect to $M$. It is clear that reducing the size
of the cycle in this way one can easily reach $C^{k-1}$ of size $2k=4$.
According to Lemma \ref{lem1}, $C^{k-1}$ is an induced cycle of length four.

The converse is evident.
\end{proof}

\begin{figure}[h]
\setlength{\unitlength}{1cm}\begin{picture}(5,1.2)\thicklines
\multiput(1,0)(1,0){3}{\circle*{0.29}}
\multiput(1,1)(1,0){3}{\circle*{0.29}}
\put(1,0){\line(1,0){2}}
\put(1,0){\line(0,1){1}}
\put(1,1){\line(1,0){2}}
\put(2,0){\line(0,1){1}}
\put(3,0){\line(0,1){1}}
\put(1.5,1.2){\makebox(0,0){$e_{1}$}}
\put(1.5,0.2){\makebox(0,0){$e_{2}$}}
\put(2.7,0.5){\makebox(0,0){$e_{3}$}}
\put(0.3,0.5){\makebox(0,0){$G_{1}$}}
\multiput(5,0)(1,0){4}{\circle*{0.29}}
\multiput(6,1)(1,0){2}{\circle*{0.29}}
\put(5,0){\line(1,0){3}}
\put(5,0){\line(1,1){1}}
\put(6,0){\line(0,1){1}}
\put(6,1){\line(1,0){1}}
\put(7,0){\line(0,1){1}}
\put(7,1){\line(1,-1){1}}
\put(5.55,0.2){\makebox(0,0){$e_{1}$}}
\put(6.5,1.2){\makebox(0,0){$e_{2}$}}
\put(7.45,0.2){\makebox(0,0){$e_{3}$}}
\put(4.3,0.5){\makebox(0,0){$G_{2}$}}
\multiput(10,0)(1,0){5}{\circle*{0.29}}
\multiput(11,1)(1,0){4}{\circle*{0.29}}
\put(10,0){\line(1,0){4}}
\put(10,0){\line(1,1){1}}
\put(11,1){\line(1,0){1}}
\put(12,0){\line(0,1){1}}
\put(13,0){\line(0,1){1}}
\put(13,1){\line(1,0){1}}
\put(13,0){\line(1,1){1}}
\put(13,1){\line(1,-1){1}}
\put(14,0){\line(0,1){1}}
\put(11.5,1.2){\makebox(0,0){$e_{1}$}}
\put(11.5,0.2){\makebox(0,0){$e_{2}$}}
\put(12.7,0.5){\makebox(0,0){$e_{3}$}}
\put(14.25,0.5){\makebox(0,0){$e_{4}$}}
\put(9.3,0.5){\makebox(0,0){$G_{3}$}}
\end{picture}\caption{$G_{1}$ is not well-coverd. $G_{2},G_{3}$ are
well-covered, but not very well-covered graphs.}%
\label{fig9}%
\end{figure}
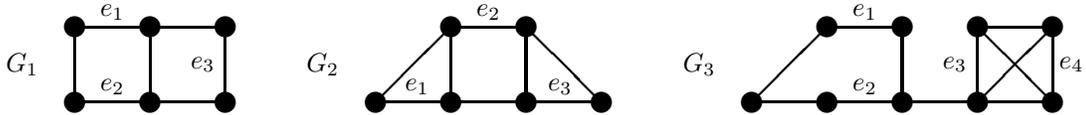

The conclusion of Lemma \ref{lem2} can be true for non-well-covered graphs;
e.g., the perfect matching $\{e_{1},e_{2},e_{3}\}$ of the graph $G_{1}$ from
Figure \ref{fig9}\ admits alternating cycles of length six and chordless of
length four. On the other hand, Lemma \ref{lem2} can fail for well-covered
graphs; e.g., the perfect matching $\{e_{1},e_{2},e_{3}\}$ of the graph
$G_{2}$ from Figure \ref{fig9} admits a unique alternating cycle of length
six, while the perfect matching $\{e_{1},e_{2},e_{3},e_{4}\}$ of the graph
$G_{3}$ from Figure \ref{fig9} admits an alternating cycle of length four that
has chords. 

\begin{theorem}
Let $G$ be a very well-covered graph. Then the following are true:

\emph{(i)} $\Psi(G)$ is a greedoid;

\emph{(ii)} $G$ has a uniquely restricted maximum matching;

\emph{(iii)} $G$ has an alternating cycle-free maximum matching;

\emph{(iv)} $G$ has an alternating $C_{4}$-free maximum matching;

\emph{(v)} every maximum matching in $G$ is alternating cycle-free;

\emph{(vi)} every maximum matching in $G$ is alternating $C_{4}$-free;

\emph{(vii)} all maximum matchings of $G$ are uniquely restricted.
\end{theorem}

\begin{proof}
Firstly, Theorem \ref{th11} implies that each maximum matching of $G$ is perfect.

\emph{(i) }$\Longrightarrow$ \emph{(ii) }Theorem \ref{th8} claims that $G$
must have a unique perfect matching, say $M$. Clearly, $M$ is a uniquely
restricted maximum matching.

\emph{(ii) }$\Longrightarrow$ \emph{(iii)} It is true, by Theorem \ref{th9}.

\emph{(iii) }$\Longrightarrow$ \emph{(iv) }Clear.

\emph{(iv) }$\Longrightarrow$\emph{ (v) }In fact, $G$ has a perfect matching,
say $M$, which is alternating $C_{4}$-free. Hence, by Lemma \ref{lem2}, $M$ is
alternating cycle-free. Consequently, by Theorem \ref{th9}, $G$ has no other
maximum matchings, and thus the assertion \emph{(v)} is true.

\emph{(v) }$\Longrightarrow$ \emph{(vi)} Clear.

\emph{(vi) }$\Longrightarrow$ \emph{(vii) }By Lemma \ref{lem2} and Theorem
\ref{th9}, it follows that every maximum matching of $G$ is uniquely restricted.

\emph{(vii) }$\Longrightarrow$ \emph{(i) }Since all maximum matchings of\emph{
}$G$\emph{ }are both perfect and uniquely restricted, it follows that $G$ has
a unique perfect matching. Consequently, $\Psi(G)$ is a greedoid, according to
Theorem \ref{th8}.
\end{proof}

\begin{corollary}
Each very well-covered $C_{4}$-free\ graph $G$ has a unique maximum matching,
and, consequently, produces a local maximum stable sets greedoid.
\end{corollary}

\begin{proof}
Combining Theorem \ref{th11} and Lemma \ref{lem2}, we infer that $G$ has a
unique perfect matching. Hence, Theorem \ref{th8} ensures that $\Psi(G)$ is a greedoid.
\end{proof}

\section{Conclusions}

In this paper we have proved that a very well-covered graph produces a local
maximum stable set greedoid if and only if it has a unique perfect matching.
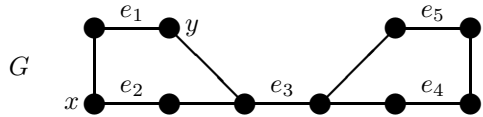
\begin{figure}[h]
\setlength{\unitlength}{1cm}\begin{picture}(5,1.2)\thicklines
\multiput(4,0)(1,0){6}{\circle*{0.29}}
\multiput(4,1)(1,0){2}{\circle*{0.29}}
\multiput(8,1)(1,0){2}{\circle*{0.29}}
\put(4,0){\line(1,0){5}}
\put(4,0){\line(0,1){1}}
\put(4,1){\line(1,0){1}}
\put(5,1){\line(1,-1){1}}
\put(7,0){\line(1,1){1}}
\put(9,0){\line(0,1){1}}
\put(8,1){\line(1,0){1}}
\put(4.5,1.2){\makebox(0,0){$e_{1}$}}
\put(4.5,0.2){\makebox(0,0){$e_{2}$}}
\put(6.5,0.2){\makebox(0,0){$e_{3}$}}
\put(8.5,0.2){\makebox(0,0){$e_{4}$}}
\put(8.5,1.2){\makebox(0,0){$e_{5}$}}
\put(3.7,0){\makebox(0,0){$x$}}
\put(5.3,1){\makebox(0,0){$y$}}
\put(3,0.5){\makebox(0,0){$G$}}
\end{picture}\caption{The well-covered graph $G$ has $\{e_{1},e_{2}%
,e_{3},e_{4},e_{5}\}$ as its unique perfect matching. $G$ is not very
well-covered, since $\alpha(G)<5=\left\vert V(G)\right\vert /2$.}%
\label{fig100}%
\end{figure}

Nevertheless, the assertion is not true for every well-covered graphs with a
unique perfect matching; e.g., $\Psi(G)$ is not a greedoid, where $G$ is the
well-covered graph from Figure \ref{fig100}, because $\{x,y\}\in\Psi(G)$,
while $\{x\},\{y\}\notin\Psi(G)$. Theorem \ref{th10} points out to a number of
examples of well-covered graphs whose families of local maximum stable graphs
form greedoids. For general well-covered graphs we propose the following.

\begin{problem}
Characterize well-covered graphs producing local maximum stable set greedoids.
\end{problem}

\end{document}